\documentclass[11pt,a4paper]{article}

\usepackage{pifont}
\usepackage[utf8]{inputenc}
\usepackage{comment}
\usepackage{url}
\usepackage{newtxtext,newtxmath}

\usepackage[utf8]{inputenc} 
\usepackage[T1]{fontenc}    
\usepackage{hyperref}       
\usepackage{url}            
\usepackage{amsfonts}       
\usepackage{nicefrac}       
\usepackage{microtype}      
\usepackage{cleveref}       
\usepackage{lipsum}         
\usepackage{graphicx}
\usepackage{doi}
\usepackage{authblk}

\begin{document}

\title{Video super-resolution for single-photon LIDAR}

\author[1,*]{Germán Mora Martín}
\author[2]{Stirling Scholes}
\author[2]{Alice Ruget}
\author[1]{Robert K. Henderson}
\author[2]{Jonathan Leach}
\author[1]{Istvan Gyongy}

\affil[1]{\small{School of Engineering, Institute for Integrated Micro and Nano Systems, The University of Edinburgh, Edinburgh, EH9 3FF, UK}}

\affil[1]{\small{School of Engineering and Physical Sciences, Heriot-Watt University, Edinburgh EH14 4AS, UK}}

\affil[*]{\url{german.mora@ed.ac.uk}}
\date{}
\maketitle



\begin{abstract} 3D Time-of-Flight (ToF) image sensors are used widely in applications such as self-driving cars, Augmented Reality (AR) and robotics. When implemented with Single-Photon Avalanche Diodes (SPADs), compact, array format sensors can be made that offer accurate depth maps over long distances, without the need for mechanical scanning. However, array sizes tend to be small, leading to low lateral resolution, which combined with low Signal-to-Noise Ratio (SNR) levels under high ambient illumination, may lead to difficulties in scene interpretation. In this paper, we use synthetic depth sequences to train a 3D Convolutional Neural Network (CNN) for denoising and upscaling ($\times$4) depth data. Experimental results, based on synthetic as well as real ToF data, are used to demonstrate the effectiveness of the scheme. With GPU acceleration, frames are processed at >30 frames per second, making the approach suitable for low-latency imaging, as required for obstacle avoidance.
\end{abstract}

\section{Introduction}

Three-Dimensional (3D) imaging captures depth information from a given scene and is used in a wide range of fields such as autonomous driving \cite{AutonomousD, vivek}, smartphones \cite{Facerec} or industrial environments \cite{Ballistics}. Time-of-Flight (ToF) is a common way to measure depth, based on illuminating the scene with a modulated light source and measuring the time for the signal to return \cite{ToF}. For longer range, outdoor applications, direct ToF sensors (dToF), and pulsed illumination are typically used, with the returning backscattered signal being detected by highly sensitive Avalanche Photo Diodes (APDs) or Single-Photon Avalanche Diodes (SPADs), and timed using an electronic stopwatch. Although the first dToF systems were based on point detectors requiring optical scanning, advances in detector technology have lead to SPAD arrays with integrated timing and processing. SPAD dToF receivers are now available in image sensor format, which, used in conjunction with flood illumination, enable high-speed 3D imaging \cite{Imagesensor1}.

\begin{figure}[h!] 
\centering\includegraphics[scale=0.36]{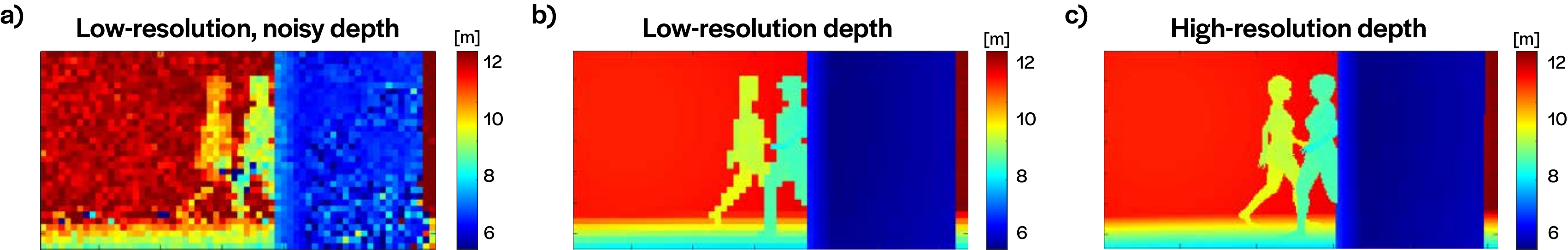}
\caption{Depth map frame of people walking using AirSim (for more details on the generation process refer to Section \ref{sec:Methods}) The frame has been produced using a 30$^{\circ}$ field-of-view and a signal-to-noise ratio of 1. a) Low-resolution depth map with noisy pixels due to solar background radiation (64$\times$32 pixels). b) Noise-free version of a), which is a nearest-neighbours resampling of the high-resolution depth map. c). High-resolution, noise-free depth map with a $\times$4 increase with respect a).}
\label{fig:LR_HR_comparison2}
\end{figure}

A drawback of current SPAD dToF image sensors is a relatively low lateral resolution \cite{SPAD}. This low resolution creates challenges in applications such as autonomous driving, where objects need to be classified as well as localised, so that appropriate action is taken \cite{DroneSense}. Although there is indication that by applying neural networks to photon timing data, the resolution problem may be overcome \cite{Turpin:20}, these approaches are yet to be generalised to arbitrary scenes. Noise in depth estimates under strong sunlight can lead to a further loss in detail in depth maps, as illustrated in Fig \ref{fig:LR_HR_comparison2}.


We therefore seek a processing scheme to improve the quality of depth maps by increasing the lateral resolution (also known as upscaling or super-resolution), whilst mitigating the effect of photon noise \cite{SRreview}. A wide-range of super-resolution methods have been proposed in literature, and although many of these are primarily devised for RGB or grayscale intensity data, they are applicable to depth too. Interpolation-based schemes such as bicubic or Lanczos resampling are computationally simple but can suffer from poor accuracy \cite{Bicubic,Lanczos}, especially when the input is noisy. Alternatively, reconstruction methods \cite{Reconstruction1,Reconstruction2} use prior knowledge from the scene such as edge smoothness to improve spatial detail at the expense of high computational costs. Learning-based methods can offer high output quality in combination with computational efficiency by learning statistical relationships between the low-resolution (LR) and the high-resolution (HR) equivalent \cite{Learning1,Learning2}. Recently, there has been increased interest in deep learning methods, which are a sub-branch of machine learning algorithms that aim to directly learn key features of the input to produce an output. These have been demonstrated to outperform previous approaches \cite{ANN,GAN,DepthSISR}.

The bulk of the research on deep-learning super-resolution methods considers the use of a single LR image to produce an HR output \cite{SRreview, gordon}. In contrast, video super-resolution schemes exploit the temporal correlations in multiple consecutive LR images to produce an improved HR reconstruction in exchange for frame rate \cite{VSRReview}. Most of these approaches are based on a combination of two networks: one for inter-frame alignment \cite{Flownet} and another one for feature extraction/fusion to produce a HR space \cite{MEMC1,MEMC2}. Other methods do not use alignment but exploit the spatio-temporal information for feature extraction using 2D or 3D convolutions \cite{3DConv}, or recurrent neural networks RCNN \cite{RCNN}. All of the aforementioned video superresolution schemes work best in the case of sub-pixel displacements between consecutive frames.

From the perspective of depth upscaling, a key limitation of existing video super-resolution schemes is that they were conceived for RGB data and thus are not optimised for the characteristics of depth frames. Although there has been some research centered on depth, the papers assume of a static scene and camera or a static scene and a moving camera  \cite{FrameAlignment,DMISR}. This paper therefore aims to devise an effective video super-resolution method targeting depth data. We consider the case of a high-speed dToF sensor \cite{HSLIDAR2}, such that displacements in consecutive frames are kept at a sub-pixel level, and account for realistic depth noise. We present a methodology for generating diverse datasets of synthetic SPAD dToF data. The resulting datasets are then used to train a super-resolution and denoising neural network. We assess the performance of the scheme using both real and synthetic input data.

\section{Generation of synthetic depth data} \label{sec:Methods}

Obtaining large and diverse datasets of noise-free, high resolution depth frames, to serve as ground truth data for super-resolution networks, is difficult in practice. In recent years, however, there have been significant advances in software packages capable of simulating realistic environments in detail. In particular, AirSim (built on Unreal Engine \cite{unrealengine}) is an open-source simulator for drones or cars with the aim of aiding AI research for autonomous vehicles \cite{Airsim}. In AirSim, one can control a vehicle through diverse virtual environment, populated with a range of objects, and retrieve virtual data from multiple sensors. The camera properties in the vehicle (position, field-of-view, lateral resolution and frame rate) can be chosen arbitrarily. In our work, RGB and depth sequences are generated at a lateral resolution of 256$\times$128 and FoV of 30$^{\circ}$ at different frame rates to ensure small object shifts between frames. A range of outdoor scenarios are simulated to provide a diverse dataset for training the neural network.

\begin{figure}[t!] 
\centering\includegraphics[scale=0.54]{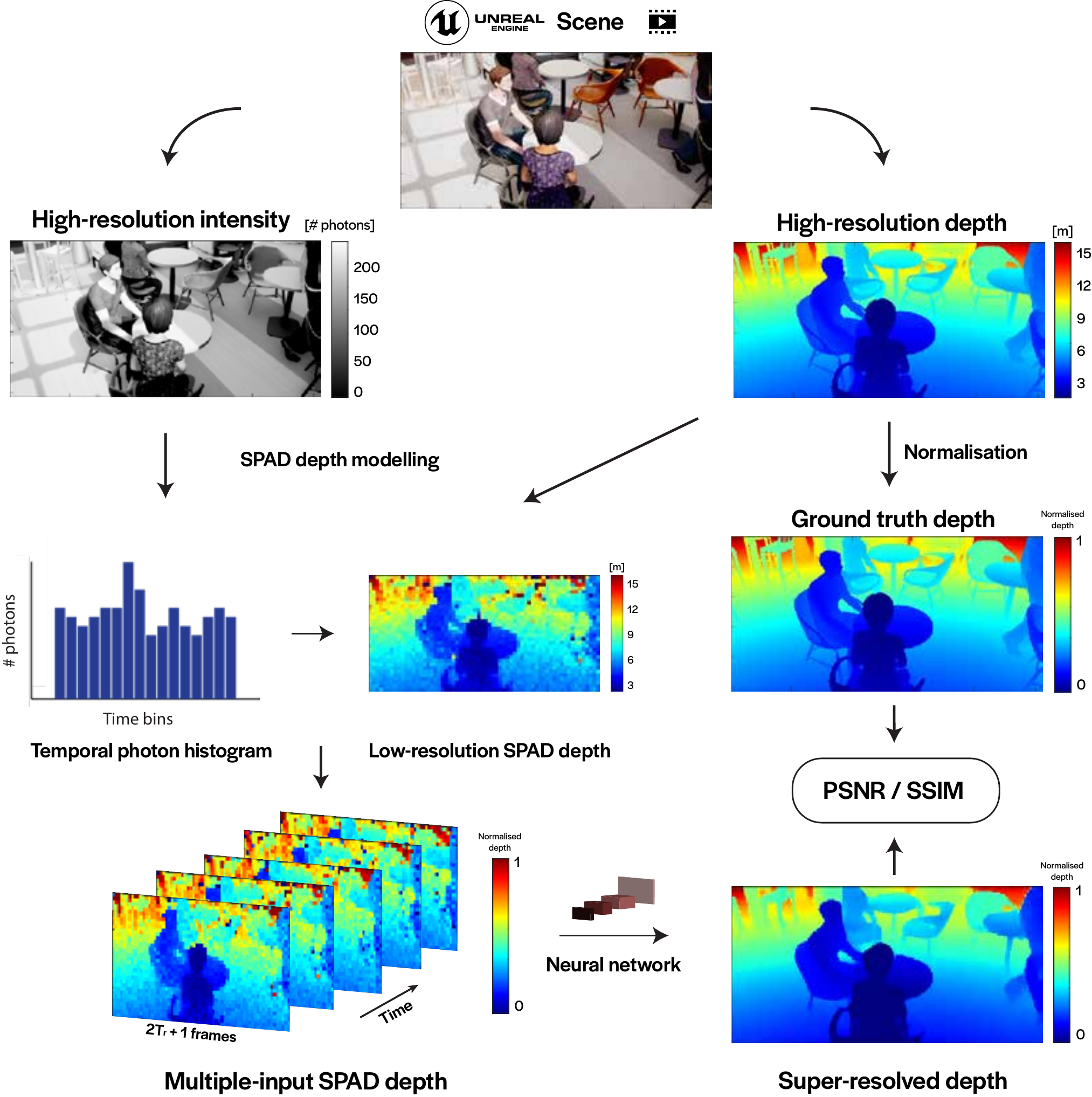}
\caption{Workflow diagram showing the process of: capturing virtual scenes; converting them into SPAD-like data; using the neural network to super-resolve frames and, analysing the network performance. It is emphasized that the captured scenes are video sequences and not single frames.}
\label{fig:Workflow_DVSR2}
\end{figure}

The depth maps provided by AirSim do not account for the main noise in single-photon dToF sensors: the inherent Poisson noise in the signal and background photon counts \cite{SPADnoise}. The high-resolution depth map of size 256$\times$128 along with the grayscale version of the high-resolution RGB frame (intensity map, 256$\times$128) are therefore used to synthesize SPAD data in the form of 16-bin temporal photon histograms (with spatial resolution 64$\times$32), where signal and background counts are set according to an optical model and randomised using the Poisson distribution. The optical model accounts for inverse square law of photon returns, and the amount of ambient illumination in the scene \cite{SPAD2}. The Signal-to-Noise Ratio (SNR) is varied to provide the network with a wide range of data, from noisy to relatively clean depth. The SNR is defined, for a given pixel, as the ratio of signal photons and background photons across all bins in the histogram. In this paper, SNR values are presented as the average in all pixels comprising a frame. To turn the histogram frames back into depth frames, centre-of-mass peak extraction is performed \cite{SPAD2}. 

The last pre-processing step is to normalise the depth data between 0 and 1 and concatenate it in groups of 2T$_R$+1 frames, where T$_R$ is the temporal radius. In other words, the input of the network consists of T$_R$ prior and posterior frames to super-resolve the central frame. The output to the neural network is compared with the ground truth depth, which is also normalised between 0 and 1, using the Peak Signal-to-Noise ratio (PSNR) and Structure Similarity Index (SSMI) \cite{PSNR,SSIM}. The metrics are calculated using the following equations:

\begin{equation} \label{Eq:PSNR}
  \mbox{PSNR} = 20 \log_{10}\left( \frac{1}{\sqrt{\frac{1}{HW}\sum_{0}^{H-1}\sum_{0}^{W-1}||Y_{GT}-Y_{SR}||^2}} \right)
\end{equation}

\begin{equation} \label{Eq:SSIM}
  \mbox{SSIM}= \frac{(2\mu_{GT}\mu_{SR}+c_1)(2\sigma_{GT-SR}+c_2)}{(\mu_{GT}^2+\mu_{SR}^2+c_1)(\sigma_{GT}^2+\sigma_{SR}^2+c_2)},
\end{equation}

\noindent where the numerator in Eq. (\ref{Eq:PSNR}) refers to the maximum value in the image (1 in our case) and the denominator corresponds to the mean-squared error. H corresponds to the image height (in pixels), W to the image width (in pixels), Y$_{GT}$ to the ground truth frame and Y$_{SR}$ to the super-resolved frame. In Eq (\ref{Eq:SSIM}), $\mu$ corresponds to the mean of the indicated frame, $\sigma$ to the variance and $c_1 = 0.0001 $ and $c_2 = 0.0009 $ are parameters to stabilize division for weak denominators \cite{3DConv}. PSNR is bound from 0 to $\infty$ and SSIM from 0 to 1. In both cases, a higher number suggests a better reconstruction. Figure \ref{fig:Workflow_DVSR2} shows the key steps in the training and assessment of the neural network proposed in this paper, from capturing scenes in AirSim, converting them into SPAD-like data and comparing the output of the network against the ground truth.

\section{Super-resolution and denoising network}

In this paper, the structure of the VSR-DUF neural network is adapted to perform video super-resolution and denoising \cite{3DConv}. The network architecture is based on blocks of 3D convolutions and a set of dynamic upsampling filters to extract spatiotemporal features without the need of performing frame realignment \cite{Flownet}. The blocks consist of batch normalisation (BN), ReLU, 1$\times$1$\times$1 convolution, BN, ReLU and 3$\times$3$\times$3 convolution and there are a total of 3+T$_R$ blocks. The network also extracts a residual map that is added to the central frame to enhance the sharpness of the final output.

The input to the network has a variable size of 64$\times$32$\times$(2T$_R$+1), depending on the temporal radius that is chosen and the output is a $\times$4 super-resolved frame in both transverse axes (256$\times$128). When the temporal radius is set to 0, single depth frame is used as input whereas a greater value (here 1 to 4) exploits the temporal correlation in multiple successive depth frames. At the start (or end) of a video sequence, there are no prior (or posterior) frames available. The corresponding frames are therefore temporally padded with a copy of the initial (or last) frame to fulfil the input size imposed by $T_R$. Additionally, the data is randomly shuffled to prevent biasing the neural network’s weights to a specific scenario. Not doing so leads to lower performance since the optimisation lands on local minima rather than the absolute minimum.

\begin{figure}[h!] 
\centering\includegraphics[scale=0.34]{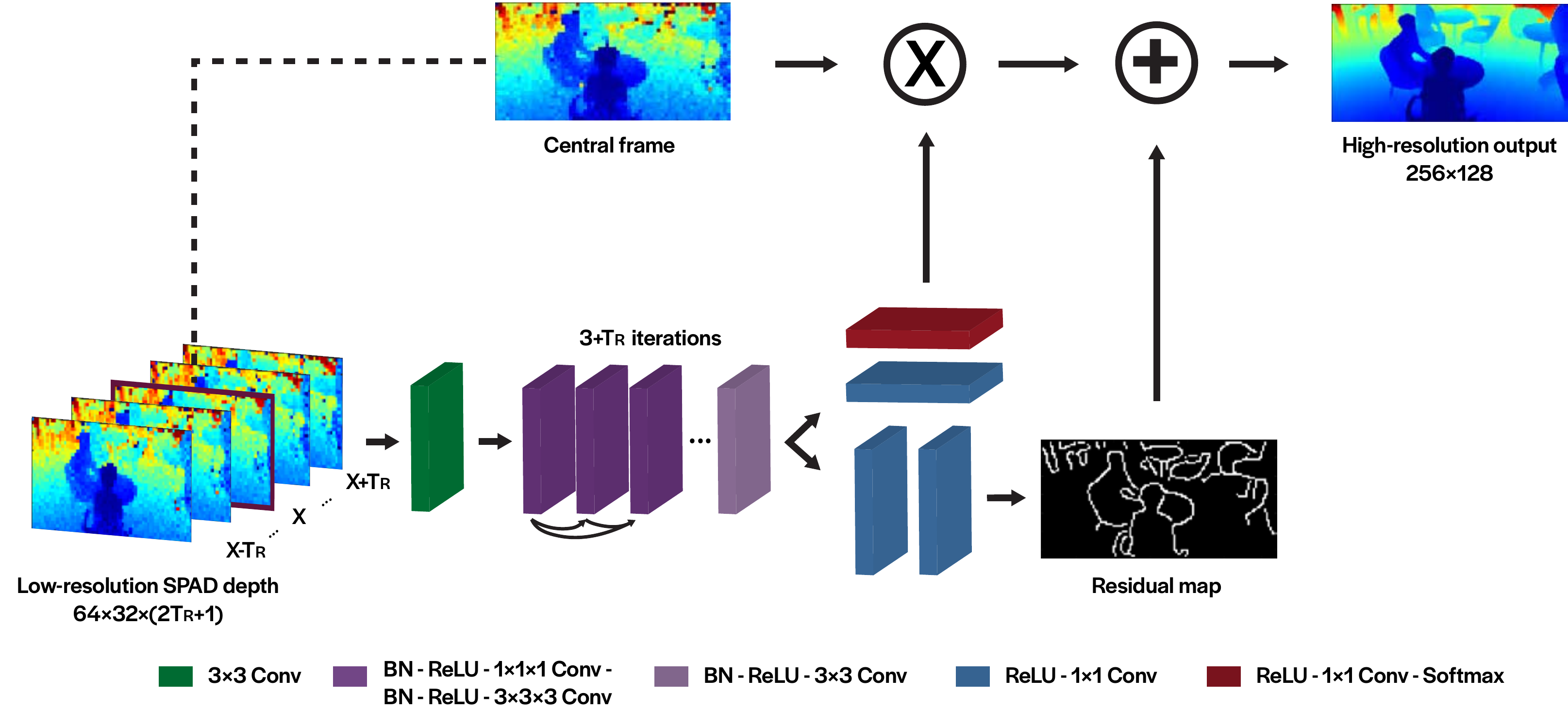}
\caption{Neural network structure, adapted from \cite{3DConv}.}
\label{fig:NN_DVSR}
\end{figure}

The neural network is implemented in Tensorflow using Keras \cite{Keras}. The training is performed using the Adam optimiser \cite{Adam} in a desktop computer (HP EliteDesk 800 G5 TWR) with the assistance of a RTX2070 GPU to accelerate the task. The loss function for this network is the Huber loss \cite{Huberloss}, often used to guarantee stable convergence during the optimisation of the model. The Huber loss is defined in Eq. (\ref{Eq:Huber}) as

\begin{equation} \label{Eq:Huber}
  \mathcal{L} = \begin{cases} \frac{1}{2}||Y_{GT}-Y_{SR}||_2^2, & ||Y_{GT}-Y_{SR}||_1 \leq \delta \\ \delta ||Y_{GT}-Y_{SR}||_1 -\frac{1}{2}\delta^2, & \mbox{otherwise} \end{cases},
\end{equation}

\noindent where $Y_{GT}$ is the HR ground truth, $Y_{SR}$ is the super-resolved HR depth map and $\delta = 0.01$ is a threshold parameter. The performance of the model is tracked at every step by the PSNR. Other parameters of interest in the neural network are the epochs and the batch size, with values of 100 and 4, respectively. Early stopping is introduced to avoid overfitting the model, thus terminating the training process after a certain amount of epochs (here a patience of 5 is used) and saving the weights corresponding to the minimum loss recorded. The learning rate of the network is set initially at 0.001 and it decreases by a factor of 10 every 10 epochs.

\section{Results}

\begin{figure}[h!] 
\centering\includegraphics[scale=0.30]{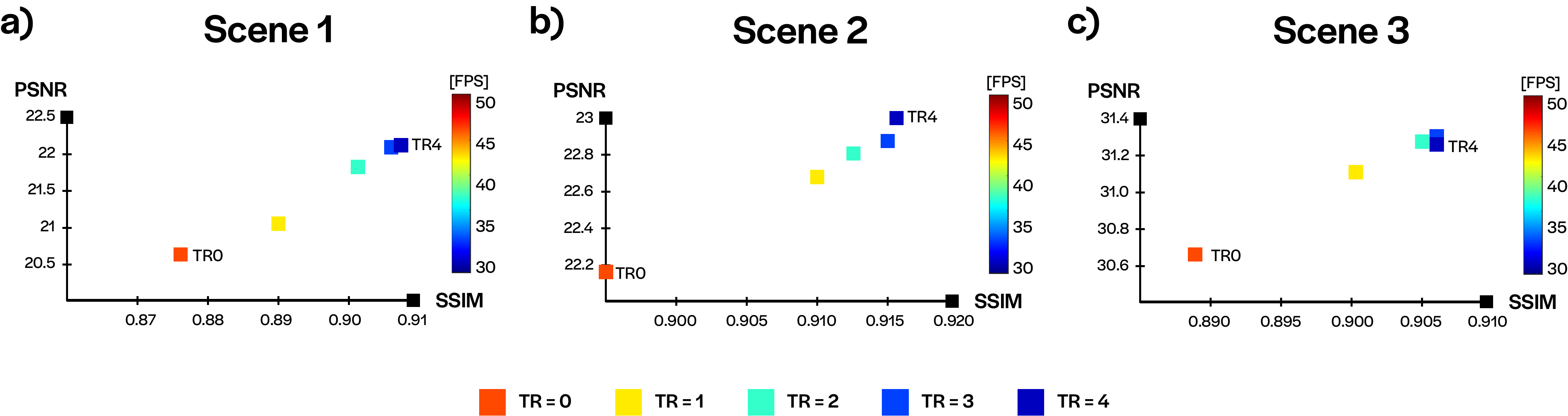}
\caption{FPS, PSNR and SSIM for different temporal radii training networks for the 3 scenes comprising the test dataset. Scene 1 and 2 have an average SNR of 1.3 and scene 3 an average SNR of 3. FPS are colour-coded and TR0 and TR4 plot points are marked for reference. }
\label{fig:Results_x4}
\end{figure}

The performance of the video super-resolution network presented here is evaluated for a model trained with 15500 examples. These include a varied range of depth maps which feature people, vehicles, bicycles, and other common objects with different noise levels (ranging from SNR 0.5 to 8), different depth ranges (ranging from 0 to 35 m) and different sampling rates, corresponding to objects shifting a variable amount of SPAD pixels in between frames (ranging from 0.2 to 5 SPAD pixels). Given the assumed FoV and SPAD pixel resolution, an object moving at 50 km/h at a distance of 35 m captured at 100 FPS, shifts by 0.5 SPAD pixels in between frames. Different versions of the network are trained for different temporal radii (ranging from 0 to 4, corresponding to 1 to 9 input images). The validation dataset consists of 1500 examples (3 sequences of 500 frames each), providing an unbiased evaluation of the neural network. Similarly, the test dataset also contains 1500 examples from 3 different sequences and is used to  perform the evaluation of the network. Table \ref{tab:Results_x4} and Fig. \ref{fig:Results_x4} summarise the performance of the network for different temporal radii in terms of PSNR, SSIM and processing speed in FPS. 

\begin{table}[htbp]
\centering
\caption{\bf Performance of super-resolution network for different temporal radii in terms of PSNR, SSIM and FPS. TR corresponds to the temporal radius. Scene 1 and 2 have an average SNR of 1.3 and scene 3 an average SNR of 3. Highest value depicted in bold. }
\resizebox{80mm}{!}{%
\begin{tabular}{cccccccc}
\hline
 & \multicolumn{2}{c}{\textbf{Scene 1}} & \multicolumn{2}{c}{\textbf{Scene 2}} & \multicolumn{2}{c}{\textbf{Scene 3}} & \textbf{} \\
 & \textbf{PSNR} & \textbf{SSIM} & \textbf{PSNR} & \textbf{SSIM} & \textbf{PSNR} & \textbf{SSIM} & \textbf{FPS} \\ \hline
\textbf{TR0} & 20.68 & \multicolumn{1}{c|}{0.877} & 22.18 & \multicolumn{1}{c|}{0.895} & 30.66 & \multicolumn{1}{c|}{0.889} & \textbf{46.9} \\
\textbf{TR1} & 21.20 & \multicolumn{1}{c|}{0.890} & 22.72 & \multicolumn{1}{c|}{0.910} & 31.17 & \multicolumn{1}{c|}{0.901} & 43.4 \\
\textbf{TR2} & 21.90 & \multicolumn{1}{c|}{0.903} & 22.82 & \multicolumn{1}{c|}{0.912} & 31.33 & \multicolumn{1}{c|}{0.905} & 38.4 \\
\textbf{TR3} & 22.04 & \multicolumn{1}{c|}{0.907} & 22.89 & \multicolumn{1}{c|}{0.915} & \textbf{31.35} & \multicolumn{1}{c|}{\textbf{0.906}} & 33.5 \\
\textbf{TR4} & \textbf{22.05} & \multicolumn{1}{c|}{\textbf{0.909}} & \textbf{23.00} & \multicolumn{1}{c|}{\textbf{0.916}} & 31.31 & \multicolumn{1}{c|}{\textbf{0.906}} & 30.8 \\ \hline
\end{tabular}%
}
\label{tab:Results_x4}
\end{table}

Visualizations 1-3 in the supplemental material show the input, the super-resolved depth maps and the ground truth for different radii for the sequences in the test dataset. Generally, the video sequences corresponding to single-image super-resolution show an improvement in lateral resolution but lack temporal coherence and introduce temporal artifacts. On the other hand, using multiple inputs reduces significantly the presence of temporal artifacts and improves the overall quality of the output frame. This is reflected by the increase of PSNR and SSIM with increasing number of inputs. The network is robust to different levels of depth noise, achieving better denoising with increasing temporal radii. In terms of computational effort, the use of multiple inputs slows the processing speed due to a larger sized network. However, even when using the maximum temporal radius considered here (TR4), the network is still able to work at video rate (30 FPS). 

The differences between TR0 and TR4 in temporal coherence are studied in more detail by considering a sequence with a person (Visualization 4). Temporal coherence (Tc) is defined here as a measure of a super-resolution method's ability to capture changes in a scene without introducing temporal artifacts. To measure it, the difference between consecutive frames (for both ground truth and super-resolved frames) is taken to obtain delta frames. All pixels with motion (according to the ground truth delta frames) are then assigned zero values. Therefore, the amount of remaining non-zero pixels gives a measure of the temporal coherence for a given frame, a larger amount indicating more pronounced temporal artifacts as quantified by Eq. \ref{Eq:Temporal_coherence}

\begin{equation} \label{Eq:Temporal_coherence}
  \mathrm{Tc}_t = \sum_{i,j=0}^{H,W}  \begin{cases} 1, & ||GT_{i,j,t}-GT_{i,j,t+1}|| = 0 \hspace{2mm} \& \hspace{2mm} ||SR_{i,j,t}-SR_{i,j,t+1}|| > 0\\ 0, & \mbox{otherwise} \end{cases}.
\end{equation}

For the sequence considered, the output for TR0 has roughly double as many pixels with artifacts compared with TR4, demonstrating that the use of multiple inputs means that motion is more accurately reproduced in the output frames. An additional study has been carried in the Supplemental material comparing the performance of this network with an upscaling factor of $\times$8. The study suggests that the $\times$8 network is not learning additional features with respect to $\times$4, but is able to produce smoother edges in certain scenarios. 

\section*{Super-resolution at different FPS}

\begin{figure}[h!] 
\centering\includegraphics[scale=0.17]{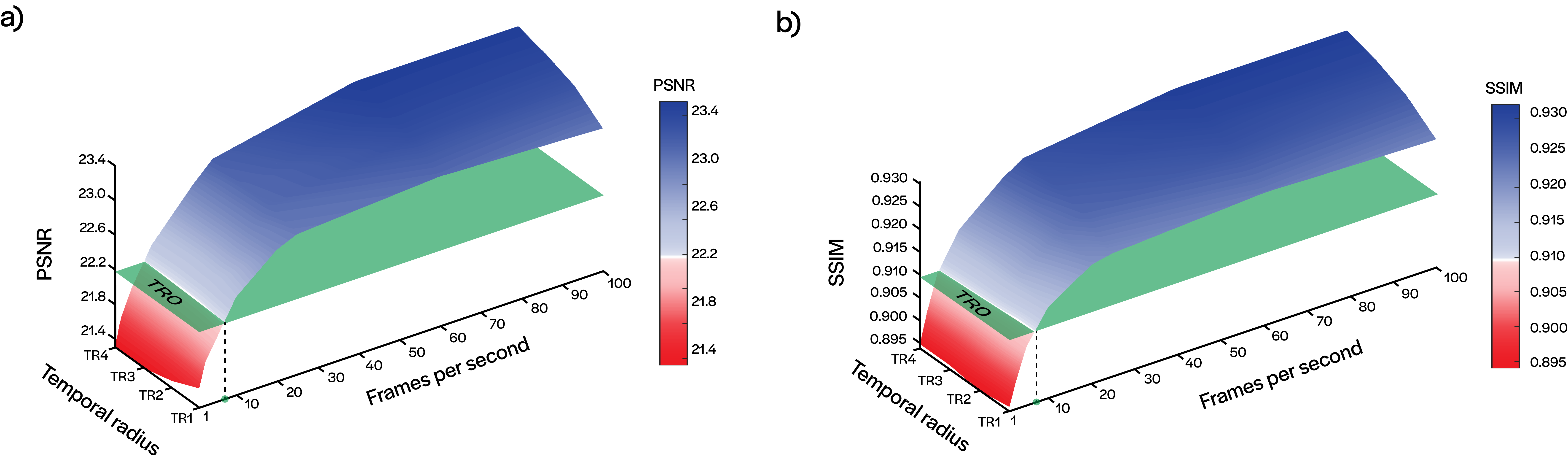}
\caption{Results of a super-resolved sequence at different temporal radii and FPS for a). PSNR. b). SSIM. The green plane denotes the average PSNR or SSIM of a single-image super-resolution technique (TR0).}
\label{fig:Speed_test}
\end{figure}

A dedicated dataset of people walking at 4.3 km/h is captured at different FPS to explore the use of temporal information in super-resolution. A main sequence is captured at 100 FPS and the lower FPS versions are generated by skipping frames with respect the original one (e.g. skip every other frame to generate a 50 FPS sequence). Given the assumed FoV and SPAD pixel resolution, people at a distance of 9 m captured at 100 FPS, shift by 0.16 SPAD pixels in between frames. As with previous studies, Table \ref{tab:Speed} shows the performance in terms of PSNR and SSIM for different temporal radii and is illustrated in Fig. \ref{fig:Speed_test}, where the parameters are compared to a single-image super-resolution approach, displayed on a green plane.

\begin{table}[htbp]
\centering
\caption{\bf Performance of super-resolution network  of a scene (SNR 1.3) in terms of PSNR and SSIM for different temporal radii and captured at different FPS. Highest value depicted in bold.}
\resizebox{100mm}{!}{%
\begin{tabular}{ccccccccccc}
\hline
 & \multicolumn{2}{c}{\textbf{TR0}} & \multicolumn{2}{c}{\textbf{TR1}} & \multicolumn{2}{c}{\textbf{TR2}} & \multicolumn{2}{c}{\textbf{TR3}} & \multicolumn{2}{c}{\textbf{TR4}} \\
\textbf{FPS} & \textbf{PSNR} & \textbf{SSIM} & \textbf{PSNR} & \textbf{SSIM} & \textbf{PSNR} & \textbf{SSIM} & \textbf{PSNR} & \textbf{SSIM} & \textbf{PSNR} & \textbf{SSIM} \\ \hline
\textbf{1} & \textbf{22.12} & \multicolumn{1}{c|}{\textbf{0.907}} & 21.52 & \multicolumn{1}{c|}{0.895} & 21.38 & \multicolumn{1}{c|}{0.894} & 21.30 & \multicolumn{1}{c|}{0.895} & 21.30 & 0.895 \\
\textbf{2.5} & \textbf{22.28} & \multicolumn{1}{c|}{\textbf{0.910}} & 21.80 & \multicolumn{1}{c|}{0.900} & 21.69 & \multicolumn{1}{c|}{0.901} & 21.66 & \multicolumn{1}{c|}{0.900} & 21.62 & 0.901 \\
\textbf{5} & \textbf{22.22} & \multicolumn{1}{c|}{\textbf{0.910}} & 22.01 & \multicolumn{1}{c|}{0.907} & 21.94 & \multicolumn{1}{c|}{0.907} & 21.91 & \multicolumn{1}{c|}{0.908} & 21.91 & 0.909 \\
\textbf{10} & 22.17 & \multicolumn{1}{c|}{0.909} & \textbf{22.43} & \multicolumn{1}{c|}{0.914} & 22.35 & \multicolumn{1}{c|}{0.916} & 22.32 & \multicolumn{1}{c|}{0.916} & 22.34 & \textbf{0.917} \\
\textbf{20} & 22.16 & \multicolumn{1}{c|}{0.909} & 22.82 & \multicolumn{1}{c|}{0.920} & 22.85 & \multicolumn{1}{c|}{0.923} & 22.86 & \multicolumn{1}{c|}{0.924} & \textbf{22.87} & \textbf{0.925} \\
\textbf{25} & 22.14 & \multicolumn{1}{c|}{0.910} & 22.92 & \multicolumn{1}{c|}{0.921} & 23.02 & \multicolumn{1}{c|}{0.925} & 23.04 & \multicolumn{1}{c|}{0.927} & \textbf{23.10} & \textbf{0.928} \\
\textbf{50} & 22.16 & \multicolumn{1}{c|}{0.910} & 23.03 & \multicolumn{1}{c|}{0.923} & 23.25 & \multicolumn{1}{c|}{0.928} & 23.36 & \multicolumn{1}{c|}{0.930} & \textbf{23.44} & \textbf{0.932} \\
\textbf{100} & 22.16 & \multicolumn{1}{c|}{0.910} & 22.95 & \multicolumn{1}{c|}{0.922} & 23.21 & \multicolumn{1}{c|}{0.927} & 23.34 & \multicolumn{1}{c|}{0.930} & \textbf{23.44} & \textbf{0.932} \\ \hline
\end{tabular}%
}
\label{tab:Speed}
\end{table}

The results indicate that when capturing at speeds greater than approximately 7 FPS, which corresponds to shifts of 2.2 SPAD pixels in between frames given the conditions above, the movement of people walking is connected sufficiently to exploit the temporal information in multiple consecutive input frames, thus the increase in performance with respect the single image approach. The general trend is that the quality of the reconstruction improves with increasing FPS, to a point where it plateaus. At higher frame rates, where the movement of objects is densely sampled, the inputs become too similar to extract additional temporal information from them. While there would not be improvement in terms of resolving features, the use of multiple inputs still has an advantage in denoising frames over single image reconstructions independently of the dynamics of the scene. It is also observed that for a given capturing speed, the performance increases with the use of more inputs, which is consistent with the results in Fig. \ref{fig:Results_x4}. Conversely, when the frame rate is low and the movement of objects is sparsely sampled, the network no longer recognises the temporal correspondence between frames. This results in lower quality reconstructions than a single image approach with the addition of temporal artifacts.

We note that the above study assumed a constant SNR. In reality, varying the frame rate can impact the exposure time, and in turn the SNR of input frames, and can lead to motion blur effects in the case of long exposures. Although not considered here, motion blur is likely to have a significant effect on the quality of reconstruction (for both single and multiple input approaches) due to the resulting disparity between the input and ground truth. We also note that this study has been carried for people, but we expect it to be generalisable to arbitrary objects in terms of the optimal shift per frame. 

\section*{Comparison with existing super-resolution techniques}

\begin{figure}[h!] 
\centering\includegraphics[scale=0.14]{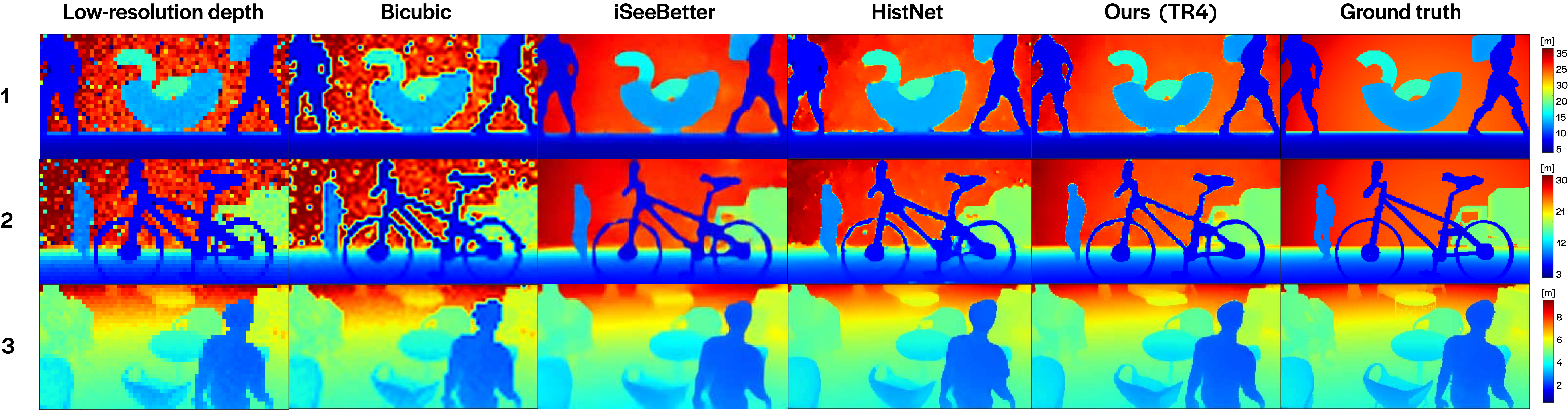}
\caption{Example output frames comparing super-resolution methods, along with the low-resolution input and ground truth.}
\label{fig:Comparison}
\end{figure}

We use the synthetic test dataset to compare the performance of our network with a bicubic interpolation method \cite{Bicubic}, a state-of-the-art neural network using SPAD histogram data (and higher resolution intensity data) \cite{Superres1} and a video super-resolution network originally for RGB images but retrained with our depth data (Visualization 5) \cite{Iseebetter}. Table \ref{tab:Comparison} and Fig. \ref{fig:Comparison}  show the performance of the different techniques in terms of average PSNR, SSIM and FPS.

\begin{table}[]
\centering
\caption{\bf Performance of super-resolution network of test dataset in terms of PSNR, SSIM and FPS for different methods (Bicubic, iSeeBetter and HistNet networks and our approach for TR1 and TR4), in bold for highest value. The network-based methods are run on a GPU. Scene 1 and 2 have an average SNR of 1.3 and scene 3 an average SNR of 3.}
\resizebox{90mm}{!}{%
\begin{tabular}{cccccccc}
\hline
 & \multicolumn{2}{c}{\textbf{Scene 1}} & \multicolumn{2}{c}{\textbf{Scene 2}} & \multicolumn{2}{c}{\textbf{Scene 3}} & \textbf{} \\
\textbf{Method} & \textbf{PSNR} & \textbf{SSIM} & \textbf{PSNR} & \textbf{SSIM} & \textbf{PSNR} & \textbf{SSIM} & \textbf{FPS} \\ \hline
\textbf{Bicubic} & 15.82 & \multicolumn{1}{c|}{0.538} & 17.60 & \multicolumn{1}{c|}{0.607} & 26.80 & \multicolumn{1}{c|}{0.841} & \textbf{185} \\
\textbf{iSeeBetter} & 20.47 & \multicolumn{1}{c|}{0.784} & 21.96 & \multicolumn{1}{c|}{0.837} & 28.74 & \multicolumn{1}{c|}{0.843} & 33 \\
\textbf{HistNet} & 19.14 & \multicolumn{1}{c|}{0.812} & 20.14 & \multicolumn{1}{c|}{0.858} & 27.18 & \multicolumn{1}{c|}{0.881} & 0.25 \\
\textbf{Ours (TR1)} & 21.20 & \multicolumn{1}{c|}{0.890} & 22.72 & \multicolumn{1}{c|}{0.910} & 31.17 & \multicolumn{1}{c|}{0.901} & 43.4 \\
\textbf{Ours (TR4)} & \textbf{22.05} & \multicolumn{1}{c|}{\textbf{0.909}} & \textbf{23.00} & \multicolumn{1}{c|}{\textbf{0.916}} & \textbf{31.31} & \multicolumn{1}{c|}{\textbf{0.906}} & 30.8 \\ \hline
\end{tabular}%
}
\label{tab:Comparison}
\end{table}

Bicubic interpolation, despite being the fastest upscaling technique, fails to remove the noise from depth data leading to poor reconstructions overall. Although PSNR can reach relatively high values, the SSIM reflects the limits of this interpolation technique. The iSeeBetter network is able to super-resolve features in a similar way with our method and removes the majority of depth noise. Additionally, with iSeeBetter, the edges of some objects (such as the bike) are reproduced in sharper way in z compared with our method (and HistNet), where these edges are somewhat averaged with the background \cite{CNNproblems}. However, on the whole, our approach appears to reconstruct objects with higher accuracy and reduced noise on surfaces. We note, for example, that objects such as the furniture (scene 3, centre) and the truck (scene 2, right) are showing less blurriness and distortion in contour and z profile than in the iSeeBetter and HistNet output. This is reflected in the higher PSNR and SSIM values for our approach compared with the iSeeBetter and HistNet. We also note, with reference to Visualization 5, that iSeeBetter and HistNet do not preserve temporal coherence, similarly to TR0 in our approach. In the case of iSeeBetter, this might be attributed to the use of an additional network for the frame re-alignment step, which is not robust to noisy images like the ones used here \cite{opticalflow}. This step can produce undesired alignments that can lead to temporal artifacts in the super-resolved frames. 

\begin{figure}[t!] 
\centering\includegraphics[scale=0.2]{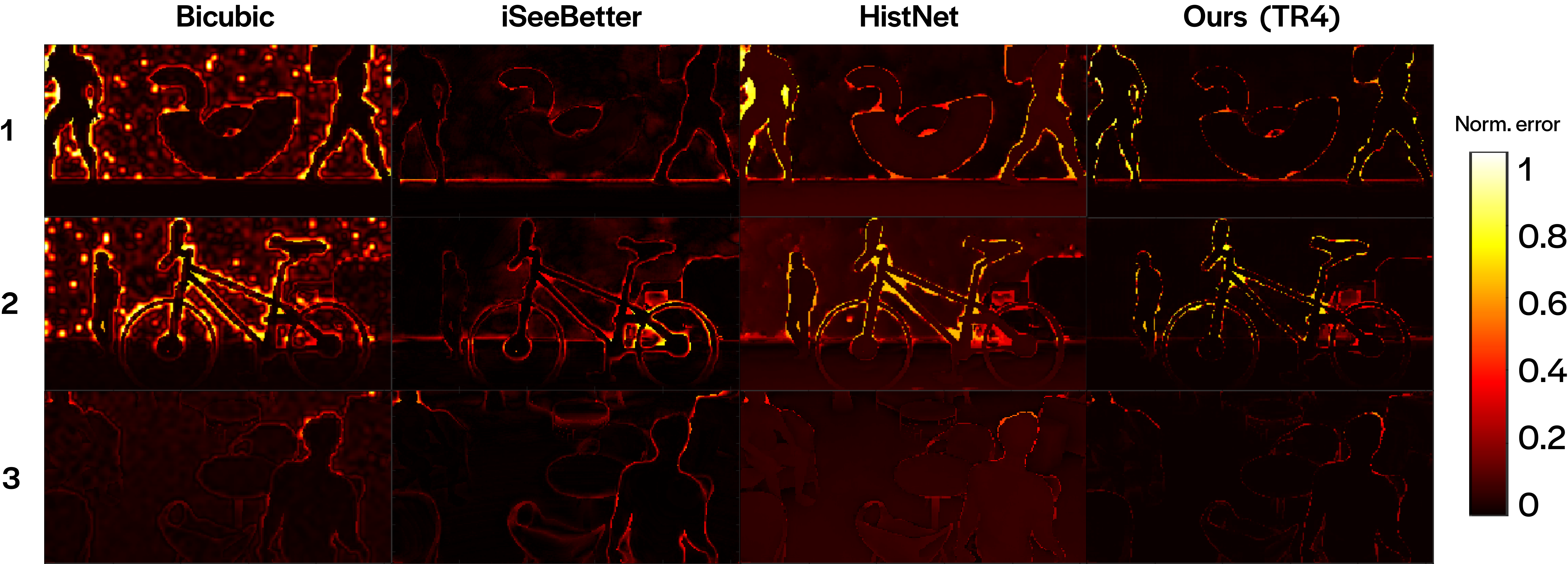}
\caption{Images indicating the error in the output of different super-resolution methods for the example data in Fig. \ref{fig:Comparison}. Depth has been normalised between 0 and 1 for displaying error images.}
\label{fig:Comparison_error}
\end{figure}

Figure \ref{fig:Comparison_error} shows the error in the reconstructions, or in other words, the absolute difference between the super-resolved frame and the ground-truth for different methods. Depths have been normalised so that error images have underlying values ranging from 0 to 1. The images provide a further illustration for the observations above. Whilst iSeeBetter frames are seen to have lower error values around some of the object edges, our method has a lower error overall across the whole frame. We observe that with our method flat surfaces are accurately reproduced whereas iSeeBetter and HistNet introduce offsets and fluctuations in depth.  

\begin{figure}[b!] 
\centering\includegraphics[scale=0.13]{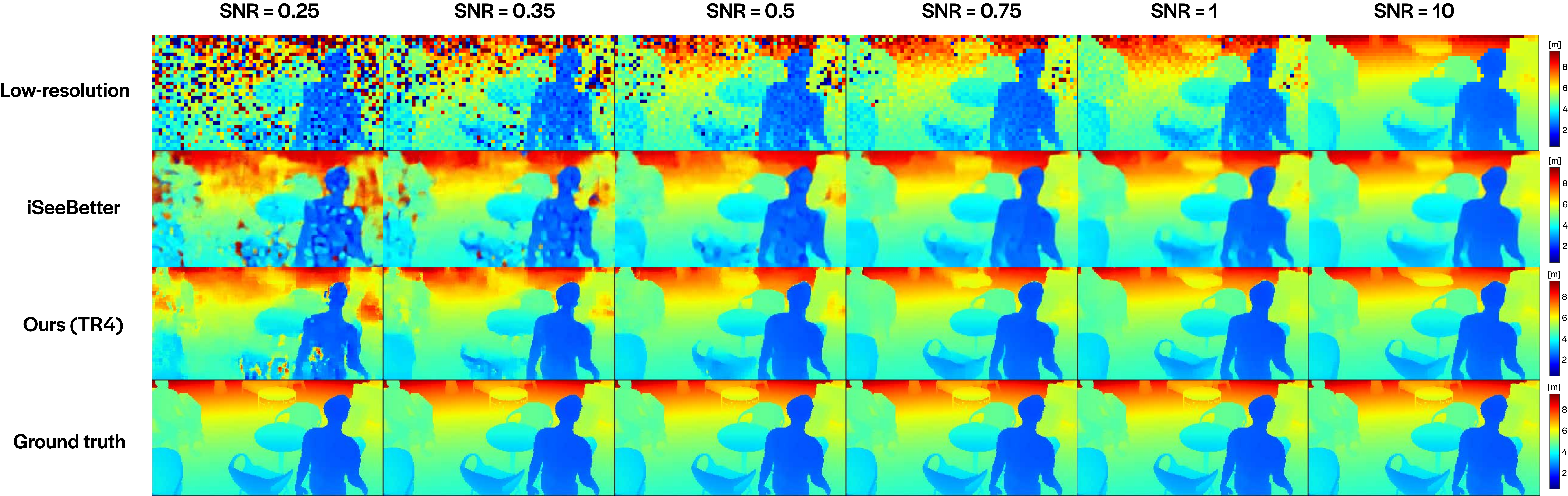}
\caption{Example frames comparing our approach with iSeeBetter for different SNR levels along with the low-resolution input and ground truth.}
\label{fig:Comparison_noise}
\end{figure}

\begin{figure}[t!] 
\centering\includegraphics[scale=0.4]{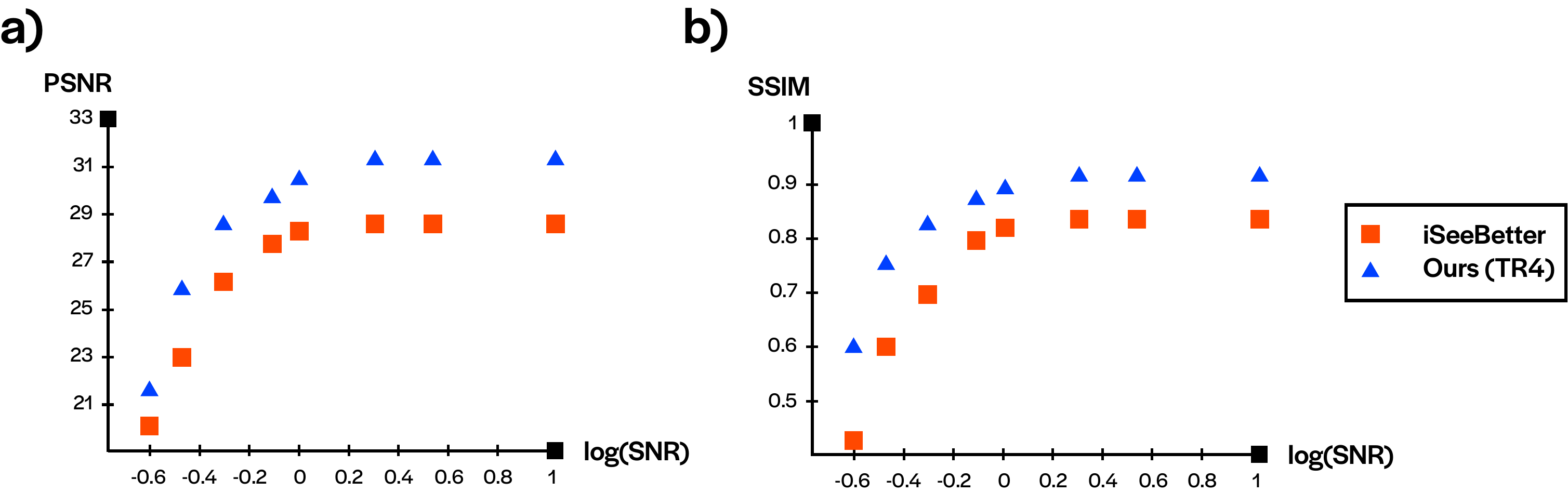}
\caption{Evolution of a) PSNR and b) SSIM metrics over SNR (in logarithmic scale) for iSeeBetter network and our approach.}
\label{fig:Test_noise_graph}
\end{figure}

The performance of the network is evaluated for a scene in the test dataset for different levels of SNR to test the robustness of the network, as compared with the iSeeBetter super-resolution network. Figure \ref{fig:Comparison_noise} shows the low resolution input for different average SNR levels (0.25, 0.34, 0.5, 0.75, 1 and 10), the super-resolved frames for iSeeBetter and our network (TR4) and the corresponding ground truth. It can be observed that for low SNR, the iSeeBetter network struggles to reconstruct the most noisy parts of the frame. On the other hand, our network appears to be more robust to noise even in the lowest SNR scenario and has overall higher PSNR and SSIM values than iSeeBetter, as can be seen in Fig. \ref{fig:Test_noise_graph}.

An additional study has been carried in the Supplemental material comparing the performance of this network with iSeeBetter as the frame rate of a synthetic depth sequence is varied. The study confirms the robustness of our approach for large shifts between frames, with the approach offering higher quality reconstructions than iSeeBetter even for inputs at low frame rates.

\section*{Super-resolution of experimental data}

\begin{figure}[h!] 
\centering\includegraphics[scale=0.175]{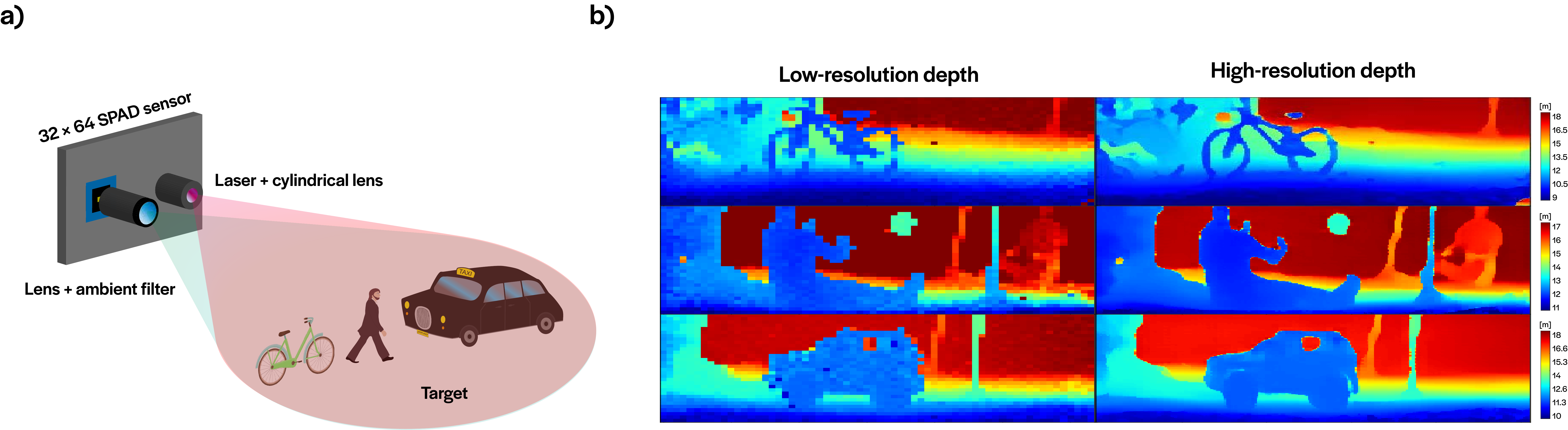}
\caption{a). Experimental setup comprising of SPAD dToF sensor, imaging lens, ambient filter, NIR laser and cylindrical lens for FoV matching. Objects at mid range (10-20 m) such as people, cars and bicycles are targeted to perform super-resolution and are shown in b). }
\label{fig:Experimental}
\end{figure}

The network is tested experimentally with a state-of-the-art SPAD dToF sensor which captures depth imaging at a pixel resolution of 64$\times$32 \cite{HSLIDAR}. The camera comprises of a NIR laser source (850 nm), which is triggered from the SPAD. The compact laser outputs a peak power of 60 W which is spread over the FoV of the SPAD (20$\times$5$^{\circ}$) by the use of a cylindrical lens to match the sensor FoV. The laser emits pulses of 10 ns with repetition rate of 1.2 MHz, suitable for mid range imaging. A 25 mm/f1.4 lens (Thorlabs MVL25M23) is used is used in front of the SPAD, together with a 10 nm bandwidth ambient filter (Thorlabs FL850-10). Figure \ref{fig:Experimental}a shows a schematic representation of the setup used during the acquisition of experimental data and Fig \ref{fig:Experimental}b depicts example depth frames, and the corresponding super-resolution output from the network

Visualization 6 shows a comparison between the input and the output of the network for different sequences captured at 50 FPS, involving a cycle, a black model car (scaled 1:4), and two people passing a ball to each other. The super-resolution frames show an improvement in the profile of objects and people, and smoother surfaces. Visualization 7 demonstrates the advantage of using multiple input frames as opposed to a single frame, in terms of resolved features, the level of denoising and temporal coherence. We note that despite the fact that the network is trained purely on virtual data, it appears to be effective when presented with real experimental data.

\section{Conclusion}

We have presented the development and application of a neural-network-based video super-resolution scheme tailored for SPAD dToF data, which tries to overcome the transverse resolution limitations of dToF image sensors. A virtual dataset involving people, cars and other common objects is gathered using Unreal and is processed into realistic SPAD data with adjustable levels of SNR. 

The results from test datasets indicate a noticeable improvement in PSNR and SSIM when using multiple inputs with respect a single image super-resolution approach. To exploit the temporal information in multiple input frames, the frame rate should be set such that object edges moving at high speed shift by less than 2-3 sensor pixelsh between frames.

A comparison study with state-of-the-art methods demonstrates the advantages of our method in offering effective reconstruction of features and denoising, whilst running at high frame rates. The network, purely trained on virtual data, also appears to visually enhance real SPAD dToF data in mid-range LIDAR.

We believe that the network could be particularly well suited to autonomous systems requiring accurate depth maps of the surroundings with low latency. These systems often rely on object detection schemes, so a potential extension of the present work would see an assessment of object detection performance when carried out on super-resolution depth frames as opposed to depth or histogram data at the native resolution of the sensor \cite{Mora-Martin:21}.

\section*{Funding}

This work was supported by EPSRC through grants EP/M01326X/1 and \newline   EP/S001638/1. Also it is supported by DSTL Dasa projects DSTLX1000147844 and DSTLX1000147352.

\section*{Acknowledgements}

The authors are grateful to STMicroelectronics for chip fabrication.

\section*{Disclosures}

The authors declare no conflicts of interest.

\section*{Supplemental Document}

See Supplement 1 for supporting content.

\section*{Data availability}

Data underlying the results presented in this paper are not publicly available at this time but may
be obtained from the authors upon reasonable request.

\bibliographystyle{ieeetr}
\bibliography{OSA-template}

\begin{thebibliography}{10}

\bibitem{AutonomousD}
X.~Chen, H.~Ma, J.~Wan, B.~Li, and T.~Xia, ``{Multi-view 3\uppercase{D} object
  detection network for autonomous driving},'' {\em Proceedings - 30th IEEE
  Conference on Computer Vision and Pattern Recognition, CVPR 2017},
  vol.~2017-Janua, pp.~6526--6534, 2017.

\bibitem{vivek}
J.~Rapp, J.~Tachella, Y.~Altmann, S.~McLaughlin, and V.~K. Goyal, ``Advances in
  single-photon lidar for autonomous vehicles: Working principles, challenges,
  and recent advances,'' {\em IEEE Signal Processing Magazine}, vol.~37, no.~4,
  pp.~62--71, 2020.

\bibitem{Facerec}
G.~Pan, S.~Han, Z.~Wu, and Y.~Wang, ``3d face recognition using mapped depth
  images,'' in {\em 2005 IEEE Computer Society Conference on Computer Vision
  and Pattern Recognition (CVPR'05) - Workshops}, pp.~175--175, 2005.

\bibitem{Ballistics}
C.~Jing, J.~Potgieter, F.~Noble, and R.~Wang, ``A comparison and analysis of
  \uppercase{RGB-D} cameras' depth performance for robotics application,'' {\em
  2017 24th International Conference on Mechatronics and Machine Vision in
  Practice, M2VIP 2017}, vol.~2017-Decem, pp.~1--6, 2017.

\bibitem{ToF}
R.~Horaud, M.~Hansard, G.~Evangelidis, and C.~Ménier, ``An overview of depth
  cameras and range scanners based on time-of-flight technologies,'' {\em
  Machine Vision and Applications}, vol.~27, p.~1005–1020, Jun 2016.

\bibitem{Imagesensor1}
R.~K. Henderson, N.~Johnston, F.~Mattioli Della~Rocca, H.~Chen, D.~Day-Uei~Li,
  G.~Hungerford, R.~Hirsch, D.~Mcloskey, P.~Yip, and D.~J.~S. Birch, ``A
  $192\times128$ time correlated \uppercase{SPAD} image sensor in 40-nm
  \uppercase{CMOS} technology,'' {\em IEEE Journal of Solid-State Circuits},
  vol.~54, no.~7, pp.~1907--1916, 2019.

\bibitem{SPAD}
S.~W. Hutchings, N.~Johnston, I.~Gyongy, T.~Al~Abbas, N.~A.~W. Dutton,
  M.~Tyler, S.~Chan, J.~Leach, and R.~K. Henderson, ``A reconfigurable
  \uppercase{3-D}-stacked \uppercase{SPAD} imager with in-pixel histogramming
  for flash \uppercase{LIDAR} or high-speed time-of-flight imaging,'' {\em IEEE
  Journal of Solid-State Circuits}, vol.~54, no.~11, pp.~2947--2956, 2019.

\bibitem{DroneSense}
S.~Scholes, A.~Ruget, G.~Mora-Martín, F.~Zhu, I.~Gyongy, and J.~Leach,
  ``Dronesense: The identification, segmentation, and orientation detection of
  drones via neural networks,'' {\em IEEE Access}, vol.~10, pp.~38154--38164,
  2022.

\bibitem{Turpin:20}
A.~Turpin, G.~Musarra, V.~Kapitany, F.~Tonolini, A.~Lyons, I.~Starshynov,
  F.~Villa, E.~Conca, F.~Fioranelli, R.~Murray-Smith, and D.~Faccio, ``Spatial
  images from temporal data,'' {\em Optica}, vol.~7, pp.~900--905, Aug 2020.

\bibitem{SRreview}
H.~Chen, X.~He, L.~Qing, Y.~Wu, C.~Ren, R.~E. Sheriff, and C.~Zhu, ``Real-world
  single image super-resolution: A brief review,'' {\em Information Fusion},
  vol.~79, pp.~124--145, 2022.

\bibitem{Bicubic}
R.~Keys, ``Cubic convolution interpolation for digital image processing,'' {\em
  IEEE Transactions on Acoustics, Speech, and Signal Processing}, vol.~29,
  no.~6, pp.~1153--1160, 1981.

\bibitem{Lanczos}
C.~E. Duchon, ``Lanczos filtering in one and two dimensions,'' {\em Journal of
  Applied Meteorology and Climatology}, vol.~18, no.~8, pp.~1016 -- 1022, 1979.

\bibitem{Reconstruction1}
S.~Dai, M.~Han, W.~Xu, Y.~Wu, Y.~Gong, and A.~K. Katsaggelos, ``Softcuts: A
  soft edge smoothness prior for color image super-resolution,'' {\em IEEE
  Transactions on Image Processing}, vol.~18, no.~5, pp.~969--981, 2009.

\bibitem{Reconstruction2}
Q.~Yan, Y.~Xu, X.~Yang, and T.~Q. Nguyen, ``Single image superresolution based
  on gradient profile sharpness,'' {\em IEEE Transactions on Image Processing},
  vol.~24, no.~10, pp.~3187--3202, 2015.

\bibitem{Learning1}
Z.~Song, Z.~Chen, and R.~Shi, ``Fast map-based super-resolution image
  reconstruction on gpu-cuda,'' in {\em Geo-Informatics in Resource Management
  and Sustainable Ecosystem} (F.~Bian and Y.~Xie, eds.), (Berlin, Heidelberg),
  pp.~170--178, Springer Berlin Heidelberg, 2015.

\bibitem{Learning2}
W.~Yang, Y.~Tian, F.~Zhou, Q.~Liao, H.~Chen, and C.~Zheng, ``Consistent coding
  scheme for single-image super-resolution via independent dictionaries,'' {\em
  IEEE Transactions on Multimedia}, vol.~18, no.~3, pp.~313--325, 2016.

\bibitem{ANN}
Y.~LeCun, B.~Boser, J.~S. Denker, D.~Henderson, R.~E. Howard, W.~Hubbard, and
  L.~D. Jackel, ``Backpropagation applied to handwritten zip code
  recognition,'' {\em Neural Computation}, vol.~1, no.~4, pp.~541--551, 1989.

\bibitem{GAN}
I.~Goodfellow, J.~Pouget-Abadie, M.~Mirza, B.~Xu, D.~Warde-Farley, S.~Ozair,
  A.~Courville, and Y.~Bengio, ``Generative adversarial nets,'' in {\em
  Advances in Neural Information Processing Systems} (Z.~Ghahramani,
  M.~Welling, C.~Cortes, N.~Lawrence, and K.~Weinberger, eds.), vol.~27, Curran
  Associates, Inc., 2014.

\bibitem{DepthSISR}
Q.~Tang, R.~Cong, R.~Sheng, L.~He, D.~Zhang, Y.~Zhao, and S.~Kwong,
  ``{BridgeNet}: A joint learning network of depth map super-resolution and
  monocular depth estimation,'' in {\em Proc. ACM MM}, 2021.

\bibitem{gordon}
Z.~Sun, D.~B. Lindell, O.~Solgaard, and G.~Wetzstein, ``Spadnet: deep rgb-spad
  sensor fusion assisted by monocular depth estimation,'' {\em Opt. Express},
  vol.~28, pp.~14948--14962, May 2020.

\bibitem{VSRReview}
H.~Liu, Z.~Ruan, P.~Zhao, C.~Dong, F.~Shang, Y.~Liu, L.~Yang, and R.~Timofte,
  ``Video super-resolution based on deep learning: a comprehensive survey,''
  {\em Artificial Intelligence Review}, 04 2022.

\bibitem{Flownet}
E.~Ilg, N.~Mayer, T.~Saikia, M.~Keuper, A.~Dosovitskiy, and T.~Brox, ``Flownet
  2.0: Evolution of optical flow estimation with deep networks,'' in {\em IEEE
  Conference on Computer Vision and Pattern Recognition (CVPR)}, Jul 2017.

\bibitem{MEMC1}
B.~Bare, B.~Yan, C.~Ma, and K.~Li, ``Real-time video super-resolution via
  motion convolution kernel estimation,'' {\em Neurocomput.}, vol.~367,
  p.~236–245, nov 2019.

\bibitem{MEMC2}
R.~Kalarot and F.~Porikli, ``Multiboot vsr: Multi-stage multi-reference
  bootstrapping for video super-resolution,'' in {\em Proceedings of the
  IEEE/CVF Conference on Computer Vision and Pattern Recognition (CVPR)
  Workshops}, June 2019.

\bibitem{3DConv}
Y.~Li, H.~Zhu, Q.~Hou, J.~Wang, and W.~Wu, ``Video super-resolution using
  multi-scale and non-local feature fusion,'' {\em Electronics}, vol.~11,
  no.~9, 2022.

\bibitem{RCNN}
X.~Zhu, Z.~Li, X.-Y. Zhang, C.~Li, Y.~Liu, and Z.~Xue, ``Residual invertible
  spatio-temporal network for video super-resolution,'' in {\em {AAAI}
  Conference on Artificial Intelligence}, 2019.

\bibitem{FrameAlignment}
G.~Mora-Martín, A.~Halimi, R.~K.~Henderson, J.~Leach, and I.~Gyongy,
  ``High-ambient, super-resolution depth imaging with a spad imager via frame
  re-alignment,'' in {\em Proceedings of IISW International Image Sensor
  Workshop}, September 2021.

\bibitem{DMISR}
J.~Kim, J.~Han, and M.~Kang, ``Multi-frame depth super-resolution for tof
  sensor with total variation regularized l1 function,'' {\em IEEE Access},
  vol.~8, pp.~165810--165826, 09 2020.

\bibitem{HSLIDAR2}
I.~Gyongy, A.~T. Erdogan, N.~A.~W. Dutton, G.~M. Martín, A.~Gorman, H.~Mai,
  F.~M. Della~Rocca, and R.~K. Henderson, ``A direct time-of-flight image
  sensor with in-pixel surface detection and dynamic vision,'' 2022.

\bibitem{unrealengine}
{Epic Games}, ``Unreal engine.''

\bibitem{Airsim}
S.~Shah, D.~Dey, C.~Lovett, and A.~Kapoor, ``Airsim: High-fidelity visual and
  physical simulation for autonomous vehicles,'' in {\em Field and Service
  Robotics} (M.~Hutter and R.~Siegwart, eds.), (Cham), pp.~621--635, Springer
  International Publishing, 2018.

\bibitem{SPADnoise}
N.~A.~W. Dutton, I.~Gyongy, L.~Parmesan, and R.~K. Henderson, ``Single photon
  counting performance and noise analysis of cmos spad-based image sensors,''
  {\em Sensors}, vol.~16, no.~7, 2016.

\bibitem{SPAD2}
I.~Gyongy, S.~W. Hutchings, A.~Halimi, M.~Tyler, S.~Chan, F.~Zhu,
  S.~McLaughlin, R.~K. Henderson, and J.~Leach, ``High-speed 3\uppercase{D}
  sensing via hybrid-mode imaging and guided upsampling,'' {\em Optica},
  vol.~7, pp.~1253--1260, Oct 2020.

\bibitem{PSNR}
F.~A. Fardo, V.~H. Conforto, F.~C. de~Oliveira, and P.~S. Rodrigues, ``A formal
  evaluation of psnr as quality measurement parameter for image segmentation
  algorithms,'' 2016.

\bibitem{SSIM}
J.~Nilsson and T.~Akenine-Möller, ``Understanding ssim,'' 2020.

\bibitem{Keras}
F.~Chollet {\em et~al.}, ``Keras,'' 2015.

\bibitem{Adam}
D.~P. Kingma and J.~Ba, ``Adam: A method for stochastic optimization,'' 2017.

\bibitem{Huberloss}
K.~Gokcesu and H.~Gokcesu, ``Generalized huber loss for robust learning and its
  efficient minimization for a robust statistics,'' 2021.

\bibitem{Superres1}
A.~Ruget, S.~McLaughlin, R.~K. Henderson, I.~Gyongy, A.~Halimi, and J.~Leach,
  ``Robust super-resolution depth imaging via a multi-feature fusion deep
  network,'' {\em Opt. Express}, vol.~29, pp.~11917--11937, Apr 2021.

\bibitem{Iseebetter}
A.~Chadha, J.~Britto, and M.~M. Roja, ``{i}{S}ee{B}etter: Spatio-temporal video
  super-resolution using recurrent generative back-projection networks,'' {\em
  Springer Journal of Computational Visual Media, September 2020, Tsinghua
  University Press}, vol.~6, no.~3, pp.~307--317, 2020.

\bibitem{CNNproblems}
X.~Song, Y.~Dai, and X.~Qin, ``Deep depth super-resolution: Learning depth
  super-resolution using deep convolutional neural network,'' in {\em Computer
  Vision -- ACCV 2016} (S.-H. Lai, V.~Lepetit, K.~Nishino, and Y.~Sato, eds.),
  (Cham), pp.~360--376, Springer International Publishing, 2017.

\bibitem{opticalflow}
P.~Ce, {\em Beyond pixels : exploring new representations and applications for
  motion analysis}.
\newblock PhD thesis, Massachusetts Institute Technology, 01 2009.

\bibitem{HSLIDAR}
I.~Gyongy, A.~T.~Erdogan, N.~A.W~Dutton, H.~Mai, F.~Mattioli Della~Rocca, and
  R.~K.~Henderson, ``A 200kfps, 256$\times$128 spad dtof sensor with peak
  tracking and smart readout,'' in {\em Proceedings of IISW International Image
  Sensor Workshop}, September 2021.

\bibitem{Mora-Martin:21}
G.~Mora-Mart\'{i}n, A.~Turpin, A.~Ruget, A.~Halimi, R.~Henderson, J.~Leach, and
  I.~Gyongy, ``High-speed object detection with a single-photon time-of-flight
  image sensor,'' {\em Opt. Express}, vol.~29, pp.~33184--33196, Oct 2021.

\end{thebibliography}

\end{document}